\documentclass[preprint2]{aastex}

\newcommand{\degree}{$^{\circ}$}
\newcommand{\arcminute}{$^{\prime}$}

\usepackage{upgreek}
\usepackage{amsmath}
\usepackage{soul}
\usepackage{color}

\begin{document}

\title{Simultaneous Observations of Giant Pulses from the Crab Pulsar,
   with the Murchison Widefield Array and Parkes Radio Telescope: Implications for the Giant Pulse Emission Mechanism.}

\author{S.I. Oronsaye\altaffilmark{1,2}, S.M. Ord\altaffilmark{1,2}, N.D.R. Bhat\altaffilmark{1,2}, S.E. Tremblay\altaffilmark{1,2}, S.J. McSweeney\altaffilmark{1}, S.J. Tingay\altaffilmark{1,2}, W. van Straten\altaffilmark{3,2}, A. Jameson\altaffilmark{3,2}, G. Bernardi\altaffilmark{4,5,6}, J.D. Bowman \altaffilmark{7}, F. Briggs\altaffilmark{8}, R.J. Cappallo\altaffilmark{9}, A.A. Deshpande\altaffilmark{10}, L.J. Greenhill\altaffilmark{5}, B.J. Hazelton\altaffilmark{11}, M. Johnston-Hollitt\altaffilmark{12}, D.L. Kaplan\altaffilmark{13}, C.J. Lonsdale\altaffilmark{9}, S.R. McWhirter\altaffilmark{9}, D.A. Mitchell\altaffilmark{14,2}, M.F. Morales\altaffilmark{11}, E. Morgan\altaffilmark{15}, D. Oberoi\altaffilmark{16}, T. Prabu\altaffilmark{10}, N. Udaya Shankar\altaffilmark{10}, K.S. Srivani\altaffilmark{10}, R. Subrahmanyan\altaffilmark{10,2}, R.B. Wayth\altaffilmark{1,2}, R.L. Webster\altaffilmark{17,2}, A. Williams\altaffilmark{1}, C.L. Williams\altaffilmark{15}}

\altaffiltext{1}{International Centre for Radio Astronomy Research (ICRAR), Curtin University, Bentley, WA 6102, Australia}
\altaffiltext{2}{ARC Centre of Excellence for All-sky Astrophysics (CAASTRO), Sydney, Australia} 
\altaffiltext{3}{Centre for Astrophysics and Supercomputing, Swinburne University, Hawthorn, Victoria 3122, Australia}
\altaffiltext{4}{Square Kilometre Array South Africa (SKASA), 3rd Floor, The Park, Park Road, Pinelands, 7405, South Africa} 
\altaffiltext{5}{Harvard-Smithsonian Center for Astrophysics, Cambridge, MA 02138, USA}
\altaffiltext{6}{Department of Physics and Electronics, Rhodes University, PO Box 94, Grahamstown, 6140, South Africa}
\altaffiltext{7}{School of Earth and Space Exploration, Arizona State University, Tempe, AZ 85287, USA} 
\altaffiltext{8}{Research School of Astronomy and Astrophysics, Australian National University, Canberra, ACT 2611, Australia} 
\altaffiltext{9}{MIT Haystack Observatory, Westford, MA 01886, USA} 
\altaffiltext{10}{Raman Research Institute, Bangalore 560080, India} 
\altaffiltext{11}{Department of Physics, University of Washington, Seattle, WA 98195, USA} 
\altaffiltext{12}{School of Chemical \& Physical Sciences, Victoria University of Wellington, Wellington 6140, New Zealand} 
\altaffiltext{13}{Department of Physics, University of Wisconsin--Milwaukee, Milwaukee, WI 53201, USA} 
\altaffiltext{14}{CSIRO Astronomy and Space Science (CASS), PO Box 76, Epping, NSW 1710, Australia} 
\altaffiltext{15}{Kavli Institute for Astrophysics and Space Research, Massachusetts Institute of Technology, Cambridge, MA 02139, USA} 
\altaffiltext{16}{National Centre for Radio Astrophysics, Tata Institute for Fundamental Research, Pune 411007, India} 
\altaffiltext{17}{School of Physics, The University of Melbourne, Parkville, VIC 3010, Australia} 
%\altaffiltext{11}{Sydney Institute for Astronomy, School of Physics, The University of Sydney, NSW 2006, Australia} 

\begin{abstract}

We report on observations of giant pulses from the Crab pulsar performed simultaneously with the Parkes radio telescope and the incoherent combination of the Murchison Widefield Array (MWA) antenna tiles. The observations were performed over a duration of approximately one hour at a center frequency of 1382 MHz with 340 MHz bandwidth at Parkes, and at a center frequency of 193 MHz with 15 MHz bandwidth at the MWA. Our analysis has led to the detection of 55 giant pulses at the MWA and 2075 at Parkes above a threshold of 3.5$\sigma$ and 6.5$\sigma$ respectively. We detected 51$\%$ of the MWA giant pulses at the Parkes radio telescope, with spectral indices in the range of $-3.6>\alpha>  -4.9$ ($S_{\rm \nu} \propto \nu^\alpha$). We present a Monte Carlo analysis supporting the conjecture that the giant pulse emission in the Crab is intrinsically broadband, the less than $100\%$ correlation being due to the relative sensitivities of the two instruments and the width of the spectral index distribution. Our observations are  consistent with the hypothesis that the spectral index of giant pulses is drawn from normal distribution of standard deviation 0.6, but with a mean that displays an evolution with frequency from -3.00 at 1382 MHz, to -2.85 at 192 MHz.

\end{abstract} 


\section{INTRODUCTION} 

The Crab pulsar (PSR B0531+21) was first revealed by its exceptionally bright pulses \citep{sta68}. These so-called ``giant pulses'' are short duration (ranging from a few ns to a few $\upmu$s) radio bursts, occurring only at the main-pulse (MP) or the inter-pulse (IP) phases of the pulsar rotation \citep{Han03,Han07,Bhat08,Karuppusamy12}. The giant pulse amplitudes and energies typically exceed those of regular pulses by two to four orders of magnitude. While the pulse energy distribution of normal pulses follows an exponential or log-normal distribution \citep{Buke-Spolaor2012}, giant pulses instead were originally observed to obey a power-law \citep{Argyle72, Hesse74,Ritchings76} suggesting that they are created by a different mechanism. However, subsequent observations \citep{cordes04} indicate that this power-law interpretation is an oversimplification only valid over small separations in radio frequency. 

The intrinsic properties of giant pulses are often obfuscated by interstellar scintillation and scattering. For instance, observations at low frequencies are significantly affected by pulse broadening that arises from multi-path scattering in the interstellar medium \citep[ISM;][]{Bhat07, Stappers11, Ellingson13}. Other than the well-known Crab, giant pulse emission has also been observed from the LMC pulsar PSR B0540-69 (J0540-6919), and a number of millisecond pulsars including PSR B1937+21 (J1939+2134) \citep{Wol84, Rom01, Knight05}. Large amplitude pulses of millisecond widths have also been observed in some long-period pulsars \citep{Ershov05}.

In the case of the Crab, giant-pulse emission is seen across the full range of the electromagnetic spectrum, from radio to gamma-rays. However, the physics governing the emission mechanism is still not well understood \citep{cordes04, Bilous11}. Inferred brightness temperatures are of the order of 10$^{30-32}$ K even when observations are affected by scattering or instrumental broadening, reaching up to $\sim$10$^{41}$ K in observations made at ultra-high time resolutions (i.e. sub-ns) \citep{Han03, Kost03, cordes04, Han07, Bhat08}. 

The short duration of each giant pulse implies broadband emission as inferred from Heisenberg-Gabor limit \citep{Gabor}. Simultaneous observations of the Crab pulsar at widely separated frequencies have been employed to investigate the validity of this assumption. \citet{Comella} first reported 50$\%$ correlation of giant pulses detected at 74 and 111 MHz. Different degrees of correlation have since been observed from as low as 3$\%$ to as high as 70$\%$ \citep{Sal99, Pop06, Bhat08, Mickaliger12}. 
%{\bf Specifically, following from the Gabor's uncertainty principle of time and frequency \citep{Gabor}, if a pulse is of very short duration, or more usually is measured by an observing system to be constrained to have arrived in a particular time interval, then the Gabor limit on the bandwidth of that signal is $df > 1/dt$. So $\sim$ns structures have GHz bandwidth.}

In order to investigate how correlated giant pulses are across wide frequency separations, we have performed an experiment using the Murchison Widefield Array (MWA) and the Parkes radio telescope. We report observations performed simultaneously with the MWA, at a frequency of 193 MHz (15 MHz bandwidth) and the Parkes radio telescope, at a frequency of 1382 MHz (340 MHz bandwidth). The MWA is a low frequency (80-300 MHz) interferometric array of 128 aperture array tiles  designed for the detection of the redshifted neutral hydrogen 21-cm signal from the Epoch of Reionization (EoR), Galactic and extragalactic surveys, Solar, heliospheric and ionospheric studies, as well as transient and pulsar studies \citep{Lonsdale09, Bowman, Tingay13}. The MWA entered into its full operational phase in mid 2013.
 
In this article, we independently  determine the time-of-arrival of pulses for each instrument. Twenty three coincident giant pulses were detected. We compare our results to those in the literature, determine an intrinsic giant pulse spectral index distribution via a Monte Carlo analysis, and discuss the constraints we can place on giant pulse emission mechanisms.

\section{OBSERVATIONS}

\subsection{Parkes observations}

The observations presented here were made  on 23 September 2013 with the 20-cm multibeam receiver system on the Parkes radio telescope. These data were recorded with a digital backend in use for the ongoing High Time Resolution Universe survey, the Berkeley-Parkes Swinburne Recorder \citep[BPSR;][]{Keith10}, from UTC 20$^{\rm h}$12$^{\rm m}$50$^{\rm s}$ to UTC 21$^{\rm h}$02$^{\rm m}$10$^{\rm s}$, with a break of 245 s to allow for system calibration (from UTC 20$^{\rm h}$39$^{\rm m}$05$^{\rm s}$ to UTC 20$^{\rm h}$43$^{\rm m}$10$^{\rm s}$). The data were recorded over a bandwidth of 340 MHz centered at 1382 MHz at each of the two linear polarisations. The polyphase filters implemented on the FPGA in the BPSR channelize the digitized data into 1024 channels with a spectral resolution of 390 kHz. The individual channels were subsequently detected and summed over both polarizations and 25 samples to yield a 64 $\upmu$s resolution time series. See Table \ref{table} for a summary of the observation parameters.

\subsection{MWA observations}

The MWA observations presented here were recorded from UTC 20$^{\rm h}$11$^{\rm m}$08$^{\rm s}$ to UTC 20$^{\rm h}$55$^{\rm m}$06$^{\rm s}$, using the MWA Voltage Capture System (VCS). The details of the VCS are presented in Tremblay et al. (2015).  

\begin{table}[h]
\begin{center}
\caption{Observation parameters. \label{table}}
\begin{tabular}{ccc}
\tableline\tableline
Parameters & MWA  & Parkes \\
\tableline
T$_{\rm sys}$ (K) & 268 & 23  \\ 
Gain (K\,Jy$^{-1}$) & 0.012$^{\rm a}$ & 0.74  \\ 
HPBW & 24{\degree} &14{\arcminute}  \\ 
Center Frequency (MHz) & 192.64 & 1382 \\
Bandwidth (MHz) & 15.36 & 340 \\
$^{\rm b}$Time Resolution ($\upmu$s) & 400 & 128 \\
Number of detected pulses & 55 & 2075 \\
Average pulse rate & \\
(pulses/minute) & 1.25 & 42.35 \\
\tableline
\end{tabular}\\
\begin{flushleft} $^{\rm a}$The net Gain is $\sim$0.042 ${\rm K\,Jy^{-1}}$ for zenith pointing.\\ $^{\rm b}$Resolution of the timeseries giant pulse search was performed on.\end{flushleft}
\end{center}
\end{table}

We used the VCS to record complex voltages in 10 kHz channels at the Nyquist-Shannon rate, resulting in one complex voltage sample every 100 $\upmu$s. The 10 kHz channels were arranged in 12 groups of 128 channels (each group representing a 1.28 MHz sub-band in the MWA system).  From each 1.28 MHz sub-band, the central 88 channels were recorded owing to a  limitation in our data recording during the commissioning of this observing mode. Voltages for both linear polarizations were recorded for 15.36 MHz (half the maximum bandwidth of the MWA system) centered at 192.64 MHz for all 128 tiles. See Table \ref{table} for a summary of the observation parameters.

\section{DATA ANALYSIS AND RESULTS}

\subsection{Parkes data analysis} \label{pda}

The data were searched in real-time for giant pulses using the {\sc Heimdall} single pulse processing software\footnote{http://sourceforge.net/projects/heimdall-astro}, following procedures similar to those described in \citet{Buke-Spolaor2011}. The giant pulse topocentric arrival times  were recorded, along with other properties including signal to noise ratio, dispersion measure (DM), and matched filter width (128 $\upmu s$). A total of 2075 giant pulses above a signal to noise ratio (S/N) of 6.5 were detected. This detection threshold is a default setting by the instrument's hardware due to the RFI environment. 

Assuming the noise is Gaussian, we can estimate the number of false positives above the detection threshold using the Gaussian distribution function. The probability, $P_n$ that we obtain an event above a threshold of $n\sigma$ is 
\begin{equation}
P_n(x>n\sigma) =\int^\infty_{\mu+n\sigma}P(x)dx = \frac{1}{2}{\rm erfc}\left[ \frac{n}{\sqrt{2}}\right],  
\end{equation}
where $\sigma$ is the rms fluctuations in the noise, $\mu$ is the mean noise level, and {\rm erfc} is the complementary error function. The number of false positives ${\it N}$, above a threshold of $\textit{n}\sigma$ is the product of the probability $P_n$, of obtaining a false positive above this threshold from one sample, and the total number of samples. Above 6.5$\sigma$, we estimated the number of false positives to be significantly less than one. Throughout the Parkes observations, we detected 3 events above $6.5 \sigma$ in the off-pulse region.  Which is higher than predicted,  and probably indicative of radio frequency interference. However the on-pulse region, to which we restrict our detections, is less than 1 percent of this extent. We therefore do not expect any false positive detections in our data set.

The flux density, \textit{S} of each giant pulse was calculated using the radiometer equation
\begin{equation}
S = \frac{(S/N)S_{sys}}{\sqrt{n_pt_{int}\Delta\nu}},
\label{radiometer}
\end{equation}
where $S_{sys}$ is the system equivalent flux density, $n_p = 2$ is the number of polarisations, $t_{int}$ = 128 $\upmu$s is integration time, and $\Delta\nu = 340$ MHz is the observing bandwidth. Given that the size of the radio telescope beam (See Table \ref{table})  is only about three times larger than the characteristic size of the Crab Nebula ($\sim$5.5\arcminute), the $S_{sys}$ is dominated by the flux contribution from the Nebula. Following \citet{Bietenholz97}, the flux contribution is $S_{CN} = 955\nu^{-0.27}$ Jy. The total system equivalent flux density is therefore
\begin{equation}
S_{sys} =  S_{syso} +  S_{CN},
\end{equation}
where $S_{syso}$ = $T_{syso}/G$ is the system equivalent flux density in the absence of the Crab Nebula, $T_{syso}$ is the system temperature in the absence of the Nebula, and \textit{G} is the gain of the telescope. This translates to a $S_{sys}$ of 906 Jy.

Although pulsars are in general stable rotators, the Crab pulsar is not. It is a young pulsar and subject to abrupt changes in rotation rate \citep{Lyne93}. Therefore a precise ephemeris, valid for the day of observation, had to be obtained in order to measure the rotation rate of the pulsar. In this case we used the time of arrival of the giant pulses at Parkes, all of which are unresolved, as markers of the pulsar rotation rate. The pulsar timing package TEMPO2 \citep{Hobbs06} was used to obtain a pulsar folding ephemeris for the observations.

\subsection{MWA data analysis}

\subsubsection{Incoherent beamforming and de-dispersion}

Post-observation, the voltage samples for the two polarizations across the 128 tiles were squared to form powers and then summed in each 10 kHz channel for each 100 $\upmu$s time step to form an incoherent beam. The sum was inverse-variance weighted to maximise the signal to noise.
This incoherent summing of the MWA tiles yields a large field-of-view, 24\degree (each tile beam $\propto \lambda/$d, where $\lambda$ is the observing wavelength, and d is the size of the tile), but a sensitivity that is 
$\sqrt{N}$ times smaller than that achievable by coherent combination of the full array, where $N$ is the number of tiles.
 
The incoherent beam was then converted into the PSRFITS data format \citep{Hotan04}. The files were then processed using the pulsar exploration and search toolkit (PRESTO; Ransom 2001). We averaged in time to 400 $\upmu$s, de-dispersed the data, and generated a time series at the nominal DM of the Crab pulsar (56.70 pc cm$^{-3}$).
\begin{figure}[h]
  \centering
  \includegraphics[scale = 0.42]{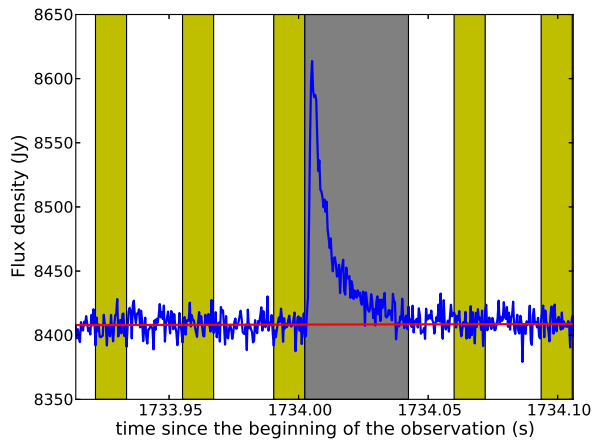}
  \caption{Dedispersed time series for the strongest giant pulse in the MWA data. The yellow (lighter) portions were used to estimate the off-pulse baseline shown by the red line. The gray (darker) area is the region that was integrated over to measure the total pulse energy. The plot spans 180 ms and has a time resolution of 400$\upmu$s. \label{pulse}}
\end{figure} 

Figure~\ref{pulse} shows the brightest giant pulse from our observations, where the scattering tail extends to $\sim$50 ms, which is longer than the pulsar rotation period of 33 ms. Further analysis of scattering and comparison with the previously published observations are discussed in \S~\ref{scatt}. 

\subsubsection{Measuring the properties of giant pulses} \label{s:sensitivity}

The Crab Nebula is an extended radio source, measuring about 5.5$^{\prime}$ in diameter, with a flux density of $\sim$955$\nu^{-0.27}$Jy \citep{Bietenholz97}, where $\nu$ is the frequency in GHz. In our case, the Nebula occupies only a tiny fraction ($\sim$0.2$\%$) of the telescope beam. The nominal sky temperature at our observing frequency and for our pointing was evaluated from the global sky model \citep{Oliveira} to be 243 K; this value explicitly includes the contribution from the Crab Nebula. At 200 MHz, our receiver temperature is 25 K \citep{Tingay13}. The incoherent gain ($G_{inco}$) of the instrument is given by the expression
\begin{equation}
G_{inco} = \frac{\frac{\lambda^2}{2\pi}(16\sqrt{N})}{2k}sin^2\theta,
\end{equation}
where $\lambda$ is the observing wavelength, \textit{N} is the number of antenna tiles, \textit{k} is the Boltzmann's constant, and $\theta$ is the elevation angle. For a nominal $G_{inco}$ of 0.012 K\,Jy$^{-1}$ at the Crab's elevation (41.42\degree), the system equivalent flux density (S$_{\rm sys}$ = T$_{\rm sys}/G_{inco}$) of our telescope is $\sim$22000 Jy. 
 
The phase of each giant pulse was determined from the arrival times of the giant pulses and the pulsar ephemeris derived from the Parkes data (see \S~\ref{pda}). We find giant pulses only at the MP and IP phases of the pulsar rotation, consistent with earlier results \citep{Lund95, cordes04, so04, Pop06, Bhat08, Karuppusamy12}. 

The total flux density was measured by integrating over a 40 ms window for every pulse period, beginning from the pulse arrival time. The integration window was determined by estimating the timescale over which the amplitude of the brightest pulse fell below the 1 $\sigma$ noise level of the baseline. This corresponded to 40 ms, approximately 6 e-folds of the estimated scattering timescale (see \S~\ref{scatt}). The baseline was determined by a linear least square fit to the off-pulse power across a series of five windows, each 12\,ms wide, distributed before and after every pulse period in the timeseries (see Figure 1). 

The resulting distribution of integrated flux density is consistent with the noise distribution of the integrated timeseries, because the number of giant pulses are few compared to the number of the observed periods ($\sim$80,000 periods). The root-mean-square noise variation (rms) was determined by estimating the standard deviation of the distribution, and the \textit{S/N} was subsequently determined. The integrated flux density of each giant pulse was then calibrated using equation \ref{radiometer}. 

In order to determine the number of false positives our pipeline generated, we simulated a noise dominated timeseries using the mean, and the variance of our data. We then subjected the simulated timeseries to the same analysis described. The number of false positives generated by this experiment above 3.5$\sigma$ was two, a false positive rate we deemed acceptable where $\sigma$ is the rms of the distribution. At this threshold, our minimum detectable flux density is $\sim$70 Jy.

In the actual analysis of the observed data-set, above a threshold of 3.5$\sigma$, 45 giant pulses were recovered from the MP phases and 10 giant pulses were recovered from the IP phase. The majority of giant pulses, 84$\%$, occur at the phase of the MP and the remaining 16$\%$ occur at the IP phase. This is in agreement with previous results published in the literature \citep{Pop07, Bhat08, Mickaliger12}.

\subsection{The giant pulse fluence distribution}

We now make the assumption, common in the literature, that the measured giant pulse energy is drawn from a power-law distribution, and attempt to determine the power-law index. We are in fact measuring the time integrated flux density, or {\it fluence}, which is a quantity directly proportional to pulse energy received by a given collecting area and will use this quantity in our analysis. 

A power-law index is often determined geometrically by a least squares fit of a straight line to logarithmically binned data \citep{Bhat08}. This is not the most reliable method of estimating this parameter and is generally considered to generate biased values of the power-law index \citep{gmy04}. In order to remove subjectivity and to limit bias we have chosen to use Hill's estimator as a maximum likelihood estimator (MLE) of the index of the underlying fluence distribution. With $n$ measurements of fluence, $x_i$, we determine the index,  $\hat{\beta}$, of the underlying power-law distribution via:
\begin{equation}
\hat{\beta} = 1 + n\left(\sum_{i=1}^{n} \ln \frac{x_{i}}{x_{\mathrm{min}}}\right)^{-1}.
\end{equation}

Hill's estimator is sensitive to the determination of the power-law cut-off ($x_{\rm min}$), but geometric  methods are also very sensitive to this and often the power-law cut-off is chosen somewhat subjectively. To limit this subjectivity we have chosen to follow a method of minimising the Kolmogorov-Smirnov distance between the fitted distribution and the measured distribution, as a function of the power-law cut-off, in order to determine the optimum value of $x_{\mathrm{min}}$. The standard error in the power-law index estimate, $\hat{\sigma} = (\hat{\beta} - 1)/\sqrt{n}$, was also determined analytically. Using this method we determined the power-law index of fluence distribution detected at the MWA to be $-$3.35 $\pm$ 0.35.

\begin{figure}[h]
	\centering
	\includegraphics[scale = 0.42]{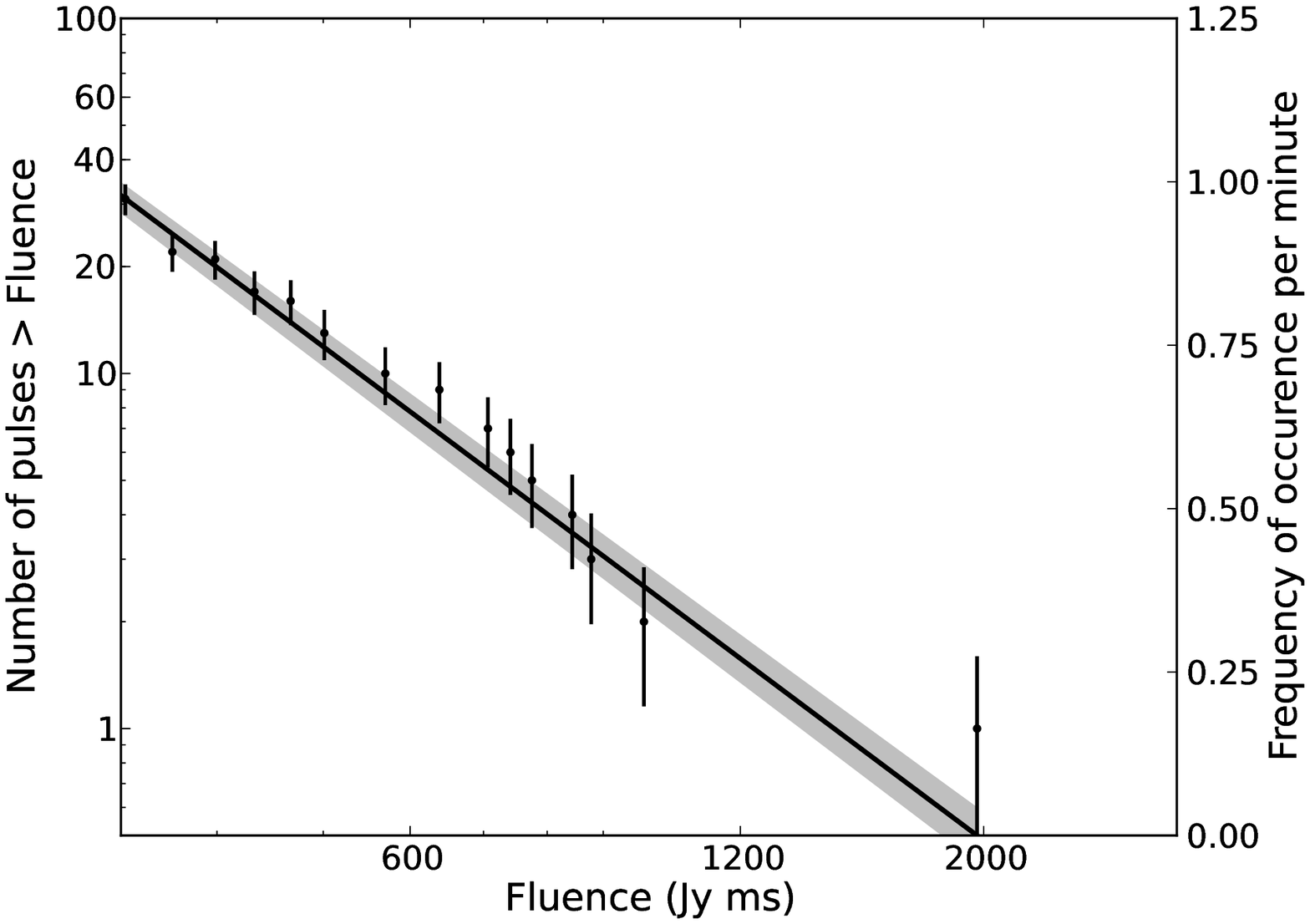}
	\includegraphics[scale = 0.42]{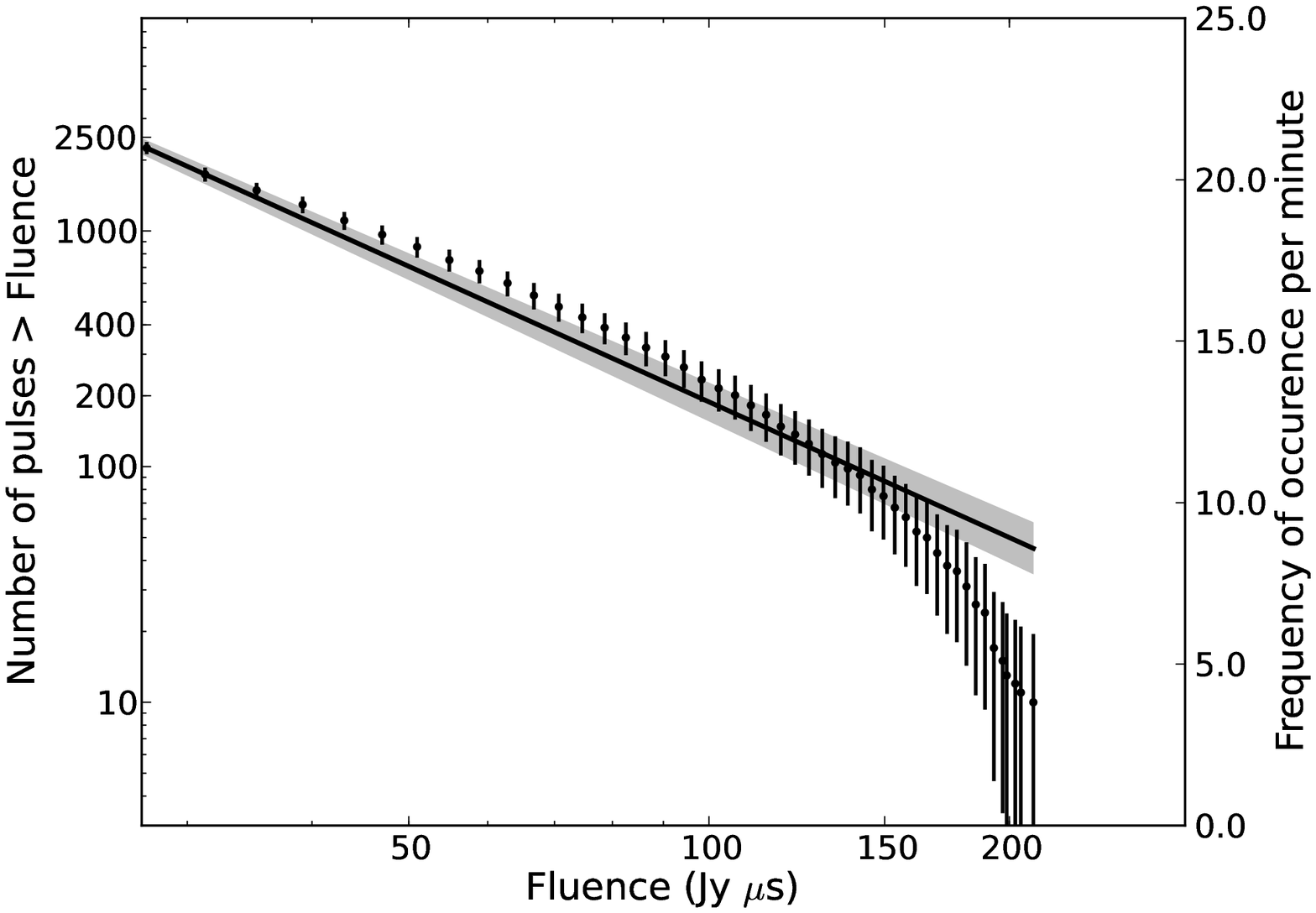}	
	\caption{Top: The cumulative distribution of the MP giant pulses detected with the MWA. Also shown here is MLE power-law fit to the distribution, with a power-law index of $-$2.35 $\pm$ 0.35. The shaded region is a $\pm0.35$ error in the power-law index. Bottom: Distribution of the MP giant pulses detected with the Parkes radio telescope. The plotted errors of fit might be underestimated. The solid line is the MLE power-law fit to the data with a power-law index of $-$2.85 $\pm$ 0.05. The shaded is region is a $\pm0.05$ error in the power-law index. \label{cdf}}
\end{figure}

The cumulative distribution (CDF) of the main-pulse phase giant pulses detected with the MWA is shown in Figure \ref{cdf}. The slope of the CDF of giant pulse fluences can also be used to determine the index of the underlying power-law distribution, although error prone, this method is often used in the pulsar literature \citep{Pop07, Karuppusamy10}. The gradient is equal to $1+\hat{\beta}$, but as the index is formed from a least squares fit of a straight line to logarithmic bins that are not independent, the error in the index is difficult to determine. In this case we find a gradient of $-2.22$ which would predict a $\hat{\beta}$ of $-3.22$, consistent with our MLE analysis.

We applied the MLE method to our observations of the Parkes pulses (at 1382\,MHz) and obtained an index of $-$2.85 $\pm$ 0.05, which is consistent with that obtained by  \citet{Karuppusamy10}; it is  slightly steeper than the $-2.33\pm0.14$ and $\-2.20\pm0.18$ at 1300 and 1470 MHz reported by \citet{Bhat08}, but is within the range of $-2.1\pm0.3$ to $-3.1\pm0.2$ as given by \citet{Mickaliger12}. 

\subsection{Pulse shape and scattering} \label{scatt}

Scattered pulse shapes are typically modelled as convolutions of intrinsic pulse shapes with the broadening functions characterising multi-path scattering through the ISM and the instrumental response. By applying a deconvolution method as described in \citet{Bhat03} to the pulse shown in Figure \ref{pulse},  we estimate a pulse broadening time, $\tau_{d} \sim 6.1 \pm 1.5$ ms, which can be compared to $\sim 0.67 \pm 0.10$ ms reported previously by \citet{Bhat07} from observations made with an early prototype built for the MWA. Our estimated $\tau_{d}$ is thus nearly 10 $\times$ larger than that measured earlier with the 3-tile prototype MWA  and 5 $\times$ larger than the extrapolated value from the observations of \citet{Karuppusamy12} at a nearby frequency of 174 MHz. However, it is a factor of two less than that expected based on recent  observations by \citet{Ellingson13} with the LWA and their re-derived frequency scaling index, $ \tau_{d} \propto \nu^{-3.7} $. As highlighted by \citet{Bhat07} and \citet{Ellingson13}, the Crab pulsar is well known for highly variable scattering, on timescales of the order of months to years, which can be attributed to ionised clouds or filaments within the nebula crossing the line of sight. 

\subsection{Coincidences in the MWA and Parkes giant pulse arrival times}

A search for coincident giant pulse arrival times was performed. Of the 55 giant pulses that were detected with the MWA, 45 were detected during the period of simultaneous observations with the Parkes radio telescope, and 1681 giant pulses were detected at Parkes. Due to the dispersive nature of the interstellar medium, the high frequency component of broadband giant pulses will arrive at the Parkes observing frequency earlier than at the MWA frequency. The Jodrell Bank monthly ephemeris\footnote{http://www.jb.man.ac.uk/pulsar/crab.html} \citep{Lyne93} provides a DM for the Crab pulsar of 56.774$\pm$0.005 pc cm$^{-3}$. The relative delay in pulse arrival time for this DM is therefore 5.746$\pm$0.008 s between 1382.00 and 200.32 MHz (upper band edge for the MWA).

\begin{figure}[h]
  \centering
  \includegraphics[scale = 0.42]{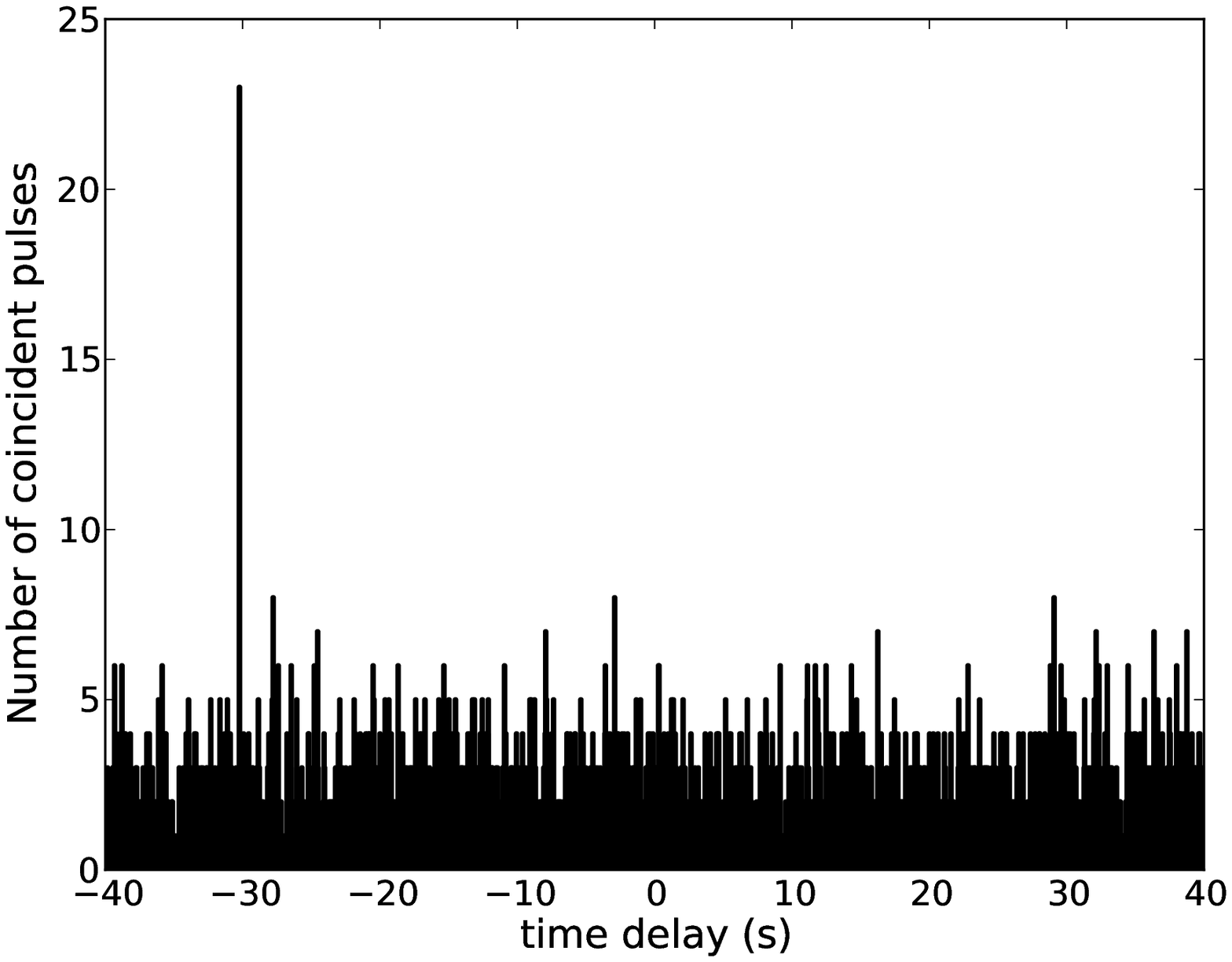}
  \caption{A plot showing the search for coincident giant pulses in the MWA and the Parkes arrival times. We identify a 36.000 s clock error in the timestamps of the recording machine used at Parkes at the time of this observation (see text). We used the arrival times of the Parkes giant pulses to search for coincidence in the MWA pulse arrival times. As can be seen in the figure, we found strong peak corresponding to 23 pulses at $-30.248$ s. Correcting for the 36 s offset yields 5.75 s which corresponds to the time delay between the arrival times of the MWA and the Parkes pulses. \label{coincidence_pulses}}

\end{figure}

In order to search for coincident giant pulses observed at the MWA and the Parkes radio telescope, we performed a correlation analysis between the arrival times of the MWA and the Parkes giant pulses. This analysis revealed that the backend used for data recording at Parkes had a clock error of 36.000 s in the timestamps at the time of this observation. As a result, the timestamps for the Parkes pulses are ahead of UTC by 36.000 s. We found 23 coincident giant pulses at $-30.248$ s (see Figure \ref{coincidence_pulses}). Correcting for the 36.000 s error yields 5.752 s delay (MWA trailing Parkes), consistent with the dispersive time delay between the arrival times of the MWA and the Parkes giant pulses. 

The 23 coincident giant pulses detected at the MWA and Parkes implies 51$\%$ correlation, since 45 giant pulses were detected with the MWA during the common observing period. Our observations are therefore consistent with previous observations (e. g. \citet{Comella, Sal99}), see \S~\ref{comparison} for comparison. 

\subsection{Spectral indices of coincident giant pulses}

\begin{figure}[h]
  \centering
  \includegraphics[scale = 0.42]{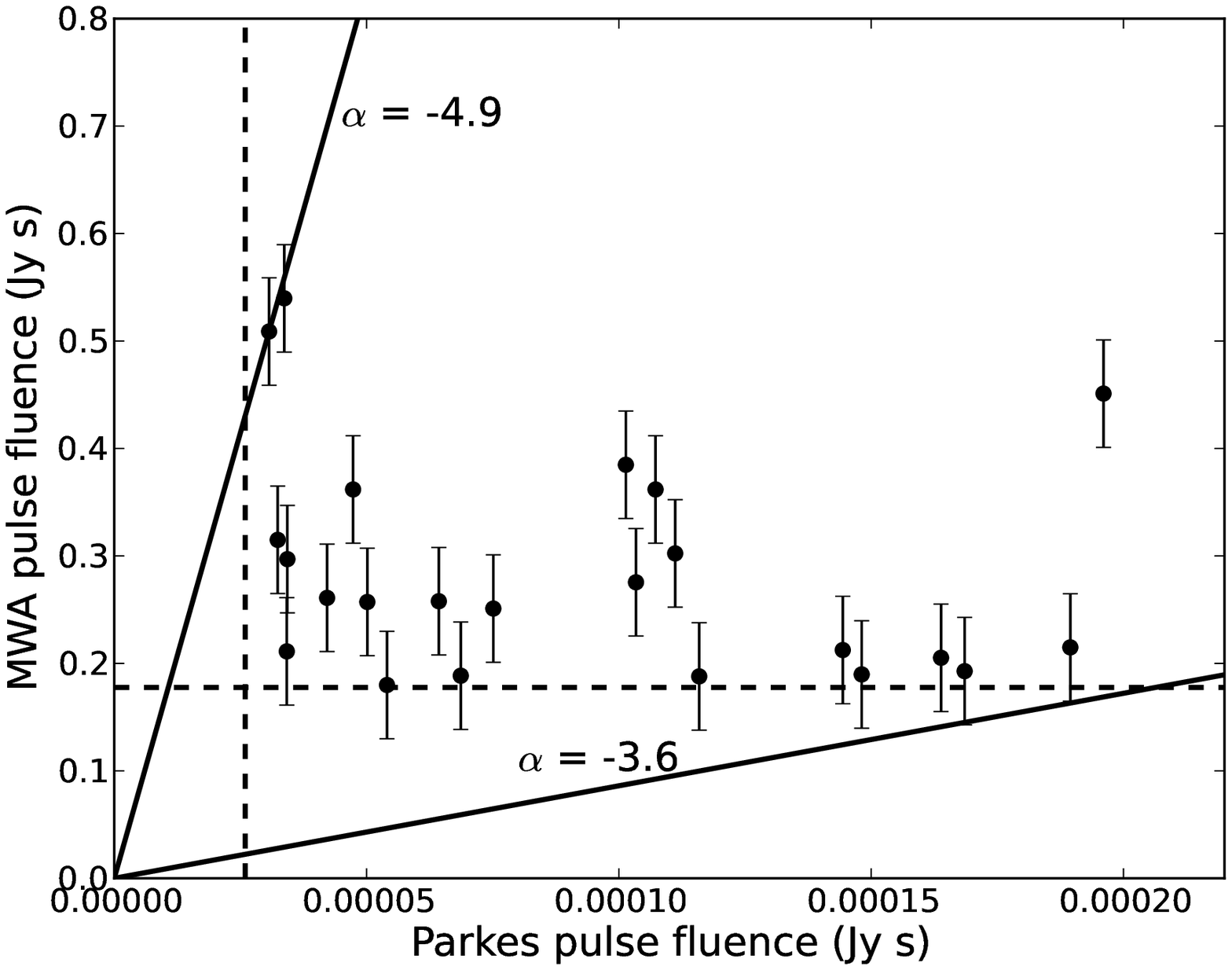}
  \caption{The fluence in Jy-s of the coincident giant pulses detected with the MWA and Parkes. The error bars indicate the uncertainty in the sky temperature measurement for the MWA (which translate to $\pm$0.08 Jy s in this case), which is negligible at the Parkes frequency. The vertical and the horizontal dash lines are the detection thresholds for the MWA and the Parkes observations. The solid lines are spectral indices $\alpha$ = -3.6 and $\alpha$ = -4.9. This figure is similar to Figure 7 of \citet{Sal99}, in which they constrained the spectral indices within the range of $ -2.2 > \alpha  > -4.9$. \label{spec_index}}
\end{figure}

The spectral indices of the giant pulses detected simultaneously at the MWA and the Parkes radio telescope were estimated using S$_\nu\propto\nu^\alpha$, where S$_\nu$ is the fluence of the giant pulse at frequency $\nu$ and $\alpha$ is the spectral index. The values derived here are presented in Figure \ref{spec_index}, and display a range between $-3.6$ and $-4.9$, which are steeper than the limits obtained by \citet{Karuppusamy10}, $-$1.44 $\pm$ 3.3 and $-$0.6 $\pm$ 3.5 for the MP and IP, respectively. Our limits are consistent with  \citet{Sal99}, based on their simultaneous detections of 29 giant pulses at 610 and 1400 MHz. \citet{Sal99} constrain the spectral index spread to be within the range $ -2.2 > \alpha  > -4.9$. The  difference in the lower limit is discussed in \S \ref{comparison}. 

\section{DISCUSSION}

\subsection{The fluence distribution}

\subsubsection{Comparison with the literature}

We detected a total of 55 giant pulses at the MWA, and estimate a power-law index of  $\beta = -3.35 \pm 0.35 $ for their pulse fluence distribution, using the maximum likelihood estimator method. For the Parkes data, in which we detected a total of 2075 pulses, application of this method yields $\beta = -2.85\pm0.05 $. In the published literature, however, the reported power-law index for the giant pulse fluence distribution covers a wide range. Different distributions have been considered, including pulse peak amplitude, average flux density, and pulse fluence, although a degree of proportionality exists between these different measures direct comparison is difficult.  Our power-law index of  $ -3.35 \pm 0.35 $ for the pulse fluence distribution at 193\,MHz is steeper than the range from  $-1.51 \pm 0.05$ to $-2.39\pm0.12$ that was reported  by \citet{Karuppusamy12} at 110-180 MHz. \citet{cordes04} estimate $-2.3$ for the index of giant pulse peak amplitude distribution at the main-pulse phase; their observation was made at 430\,MHz and they note that the distribution only obeys a power-law for a limited range of pulse flux density. However our estimated $\beta$ is in good agreement with a value of $  -3.46 \pm 0.04 $ determined by \citet{Lund95} for the average flux density distribution from their extensive observations at 800\,MHz.

We propose that the majority of the published variation in power-law indices arises from the underlying distribution not following a simple power law over a large range of fluences and also from inconsistencies in the subjective determination of the power-law cut-off. The MLE based method that we used here has the virtue of removing some subjectivity from this analysis.

\subsubsection{Effects of scintillation}

Observed radio pulses are often influenced by interstellar scintillation \citep{Lyne68, cordes04}. Under unfavorable conditions, this effect can strongly modulate the observed fluence of giant pulses. In our case, this effect may arise from refractive modulation since short-term  diffractive modulations will be highly quenched over
our observing duration and bandwidth, at both MWA and Parkes frequencies. \citet{cordes04} report $\Delta \nu_d < 0.8$ MHz and $\Delta t_d = 25 \pm 5$ secs at 1.48 GHz; where $\Delta \nu_d$ and $\Delta t_d$ are the diffractive scintillation bandwidth and timescale respectively. At the MWA frequency these will be orders of magnitude smaller and are therefore no longer relevant. As for refractive modulation, \citet{Rickett90} report a time scale of $\sim83 \pm 50$ days at 196 MHz and $\sim6 - 12$ days at their highest frequency of 610 MHz.  Even if we consider a factor of two variability as a possible worst case scenario at our frequencies, it may not still significantly alter our spectral index analysis and conclusions.

%Observed radio pulses are often influenced by interstellar scintillation \citep{Lyne68, cordes04}. Under unfavorable conditions, this effect can strongly modulate the observed fluence of giant pulses. In our case, this effect may arise from refractive modulation since short-term or diffractive modulations will be highly minimized at both MWA and Parkes frequencies. For instance, \citet{cordes04} shows that $\Delta \nu_d < 0.8$ MHz and $\Delta t_d = 25 \pm 5$ secs at 1.48 GHz; where $\Delta \nu_d$ and $\Delta t_d$ are the diffractive scintillation bandwidth and timescale respectively. At the MWA frequency they will be orders of magnitude smaller and are no longer relevant. For refractive modulation, \citet{Rickett90} reports a decrease in the level of refractive variability with decreasing frequency; they reports a time scale of $\sim83 \pm 50$ days at 196 MHz and $\sim6 - 12$ days at their highest frequency. If we consider a factor of 2 for a worst case scenario across our frequencies, this will not significantly modulate the observed fluence of giant pulses, as well as the spectral index analysis.}

\subsection{Correlated detections and the intrinsic giant pulse spectral index distribution} \label{comparison}

Simultaneous observations of the Crab pulsar at multiple frequencies have been reported in the literature. This includes the work of \citet{Comella} from early pulsar observations, at 74 and 111 MHz, to observations that reach frequencies as high as 8.9 GHz \citep{Mickaliger12}. The level of correlation presented does vary. Observations by \citet{Ellingson13} using the LWA with 4 x 16 MHz observing bands spread from 28 MHz to 76 MHz contain 33 giant pulse detections, only one of which was observed in all 4 bands, but almost all were observed in at least 2. Based on observations with the Green Bank 25-metre telescope and the Very Large Array, \citet{Sal99} reported the detection of 70$\%$ of giant pulses seen simultaneously at both 600 and 1400 MHz.  \citet{Pop06} report 27$\%$ simultaneous detections at their observing frequencies of 23 and 600 MHz but only 16$\%$ at 111 and 600 MHz. \citet{Bhat08} found 70$\%$ correlation between the the giant pulses observed at 1300 MHz and 1470 MHz, similar to the result obtained by \citet{Sal99}. The observations of \citet{Mickaliger12} at larger frequency separations show a significantly smaller fraction, 3$\%$ to 5$\%$ between 1.2 GHz and 8.9 GHz for the MP and IP giant pulses, respectively. 

Our observations display a correlation of 51\% between the pulses detected at Parkes and at the MWA, but the range of spectral indices displayed by the detected pulses ($-3.6$ to $-4.9$) is considerably narrower than reported by \citep{Sal99}. It is likely that the paucity of pulses with spectral indices shallower than $-3.6$ is due to the sensitivity limit of the MWA telescope and consequently that the intrinsic degree of correlation is likely to be much higher than 50\%. We hypothesise that the apparent reduction of correlation is due to the wide range of spectral index displayed by the giant pulses, coupled with the sensitivity limits of the two telescopes.

In order to investigate this further, a Monte Carlo simulation was conducted using the Parkes data as the known distribution of giant pulse fluence. Attempts were made to predict the number of pulses detected coincidentally at the MWA and their spectral index distribution using a two sample Komolgorov-Smirnov (KS) test. 

We first determined the most likely intrinsic distribution of giant pulse spectral indices.  As reported by both \cite{Sal99} and Karuppusamy et al. (2010), the scatter in individual giant pulse spectral indices is large. The hypothesis that the intrinsic giant pulse spectral indices could have been drawn from a uniform distribution between $-2.2$ and $-4.9$ was immediately discounted; it reproduced the range of detected spectral indices ($-3.6$ to $-4.9$) however it over-estimated the number of coincident detections by a factor of 20. The broad distributions favoured by  Karuppusamy et al. (2010)  were also discounted by this analysis, as although they could reproduce the number of coincident detections, the distribution of predicted spectral indices were not compatible with the observations. A normal distribution of spectral indices was a much more successful hypothesis. It simultaneously satisfied the constraints of the detected spectral indices and the number of detections. We chose to investigate this further and determine the parameters of such a distribution that would best fit our data. We allowed the mean of the distribution to vary between $-2.0$ and $-3.6$, but held the standard deviation to 0.6, consistent with the observations of \cite{Sal99}, and investigated the predicted number of coincident detections and their spectral indices. 

For each of the Parkes giant-pulse detections we randomly drew a spectral index from the distribution under test and predicted the fluence of the pulse at the MWA. If the predicted fluence was above the MWA detection threshold we registered a detection and the spectral index that was associated with it. This was repeated for every pulse. Each trial distribution was tested 1000 times and the results combined.
\begin{figure}
	\centering
	\includegraphics[scale = 0.42]{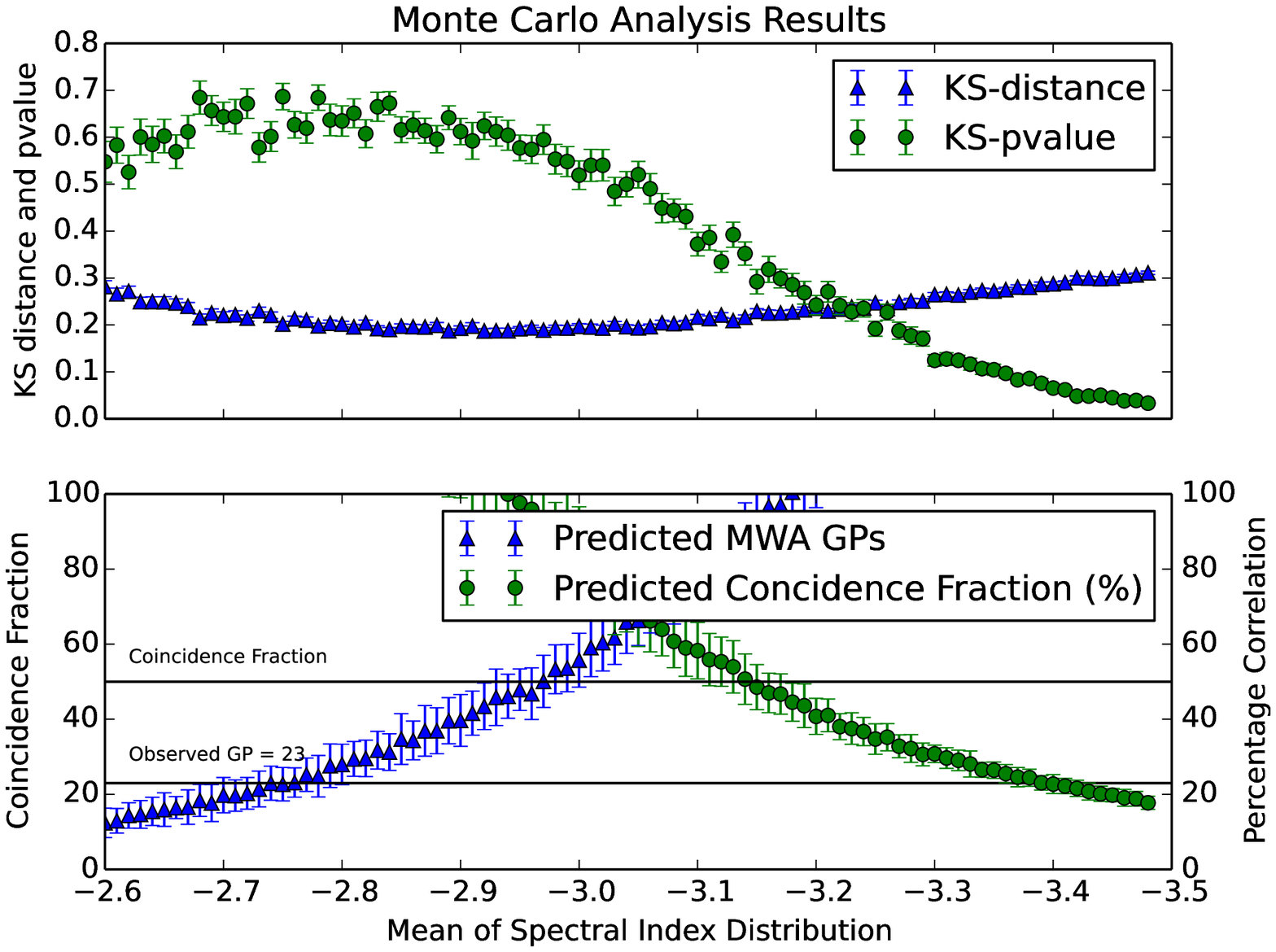}		
	\caption{The results of the Monte Carlo analysis to determine the best fit intrinsic mean spectral index of the giant pulses. The three constraints, number of coincident detections (23), observed correlation factor (50\%) and distribution of observed spectral indices cannot be satisfied by the same input distribution. \label{no-fine-tuning}}
\end{figure}

As we have recorded observations at both telescopes we can reverse this analysis and use the MWA detected pulse and spectral index distribution to predict the coincident pulses at Parkes, and thereby determine the percentage of correlated pulses. We are only observing the steep spectrum end of the giant pulse distribution at the MWA, as evidenced by inspecting the measured spectral index distribution. We can infer that some pulses detected at the MWA will have a combination of fluence and steep spectral index that would bring them below the detection threshold at  Parkes. This feature would be the reason for the less than 100\% correlation in pulse detections.

In Figure \ref{no-fine-tuning}, the lower panel shows the predicted number of GPs detected at the MWA for each trial spectral index distribution as triangles and the percentage of coincident GPs between the MWA and Parkes. There are also two lines representing the observed number of GPs and the observed coincidence fraction. To satisfy the observations, a trial spectral index distribution should simultaneously satisfy both these constraints. The error bar is the standard deviation about this average prediction. The upper panel shows the results, both distance and p-value, for the two-sample KS test between the predicted spectral index distribution of the coincident pulses and the measured distribution; the error bar is the standard error. The p-value can be interpreted as the probability that the distance is as large as observed, if the null hypothesis is true (being that both populations are drawn from the same distribution). The null hypothesis is typically not rejected if the distance is below a critical value for the two distributions under test. These results show that, for those distributions with a significant p-value, we cannot reject the null hypothesis and that the observed and simulated spectral index distributions are consistent with same parent distribution.

Without any fine tuning, the simulations predict the following:

\begin{itemize}
\item{The detected spectral index distribution at the MWA is best replicated by an input spectral index distribution of -2.85 with a standard deviation of 0.6}
\item{The observed 50\% correlation is best replicated by a spectral index distribution closer to -3.15}
\item{Both these input distributions over-estimate the number of coincident pulses by a factor of 2 to 3}
\end{itemize}

\subsubsection{Fine tuning the analysis}

The simplest, and most reasonable fine tuning is to reflect the lack of flux density calibration and assume some less-than-theoretical detection sensitivity. We can align the number of MWA detections with the best fit spectral index distribution, using the MWA detection threshold as a fit parameter. Raising the MWA detection threshold by 10\% generates a best fit mean spectral index distribution of -2.8, and correctly predicts the number of MWA detections. However it cannot simultaneously replicate the correlation percentage, which is now reproduced by invoking a source spectral index distribution with a mean of -3.2.

Altering the width of the model spectral index distribution does not provide the required fine tuning. The width of 1.2 is wide enough that it produces features similar to that generated by a uniform distribution of spectral indices. In that we can reproduce the number of detections at the MWA but the distribution of “observed” spectral indices differs too greatly. In the case of the narrow distribution, it is difficult to reproduce the number of detections unless we adjust the relative sensitivity of the telescopes beyond what would be considered reasonable. 

A distribution that satisfies the constraints in predicting the Parkes pulse survival to the MWA observing frequency is therefore still inconsistent with the distribution that satisfies the degree of correlation. A more sophisticated fine-tuning would be to invoke some evolution in the distribution as a function of frequency. We have found that this evolution does not have to be extreme in order to remove  the inconsistency.  A 5\%  flattening  in the spectral index, on average, between the two frequencies is all that is required to match the observed pulse coincidence rate, while maintaining a spectral index distribution consistent with that observed.  We obtained this estimate using the previous Monte Carlo analysis, but including an average reduction factor in the randomly drawn individual pulse spectral index when predicting the observed fluence at the MWA. We incorporated the same factor in the inverse experiment.

\begin{figure}
	\centering
	\includegraphics[scale = 0.42]{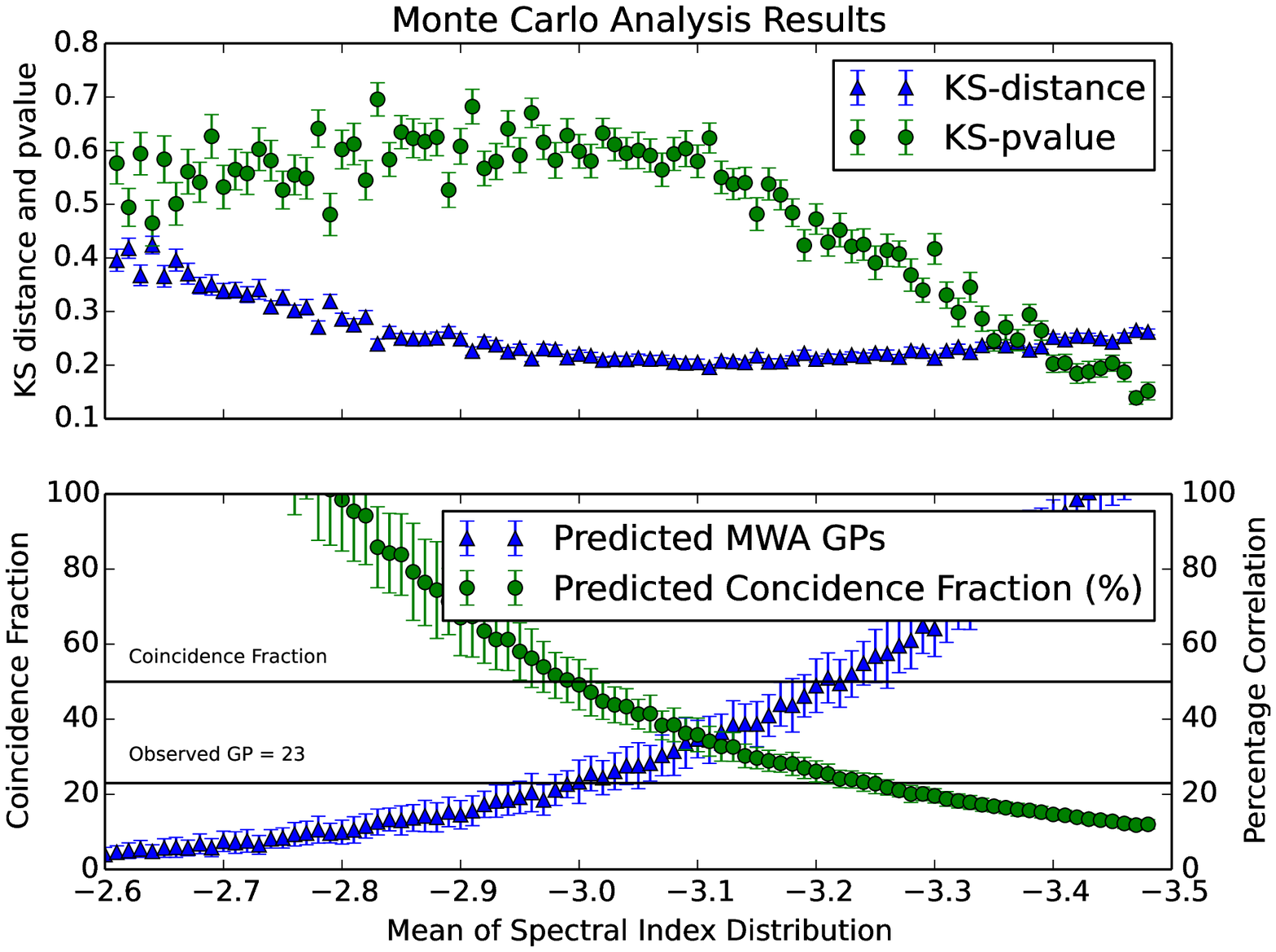}	
	\caption{Fine tuning the Monte Carlo analysis by incorporating an overestimate of the MWA sensitivity by 10\% and a mild (5\%) evolution in the spectral index as a function of frequency produces an internally consistent estimate for the mean of the spectral index distribution($-3.00$ at 1382 MHz, to greater than $-2.85$ at 192 MHz). \label{fine-tuning}}
\end{figure}

Incorporating this fine tuning, our observations are  consistent with the hypothesis that the spectral index of giant pulses is drawn from normal distribution of standard deviation 0.6, but with a mean that displays an evolution with frequency from -3.00 at 1382 MHz, to -2.85 at 192 MHz. Note this estimation only includes the integrated effect of the evolution in spectral index, and therefore does not incorporate curvature. It would be consistent with these results if the low frequency spectral index were considerably flatter, or the high frequency steeper, than that estimated here, but that would require a rapid evolution, such as a turn-over, to be present. Figure \ref{fine-tuning} shows the results of the analysis including the fine tuning and we summarise the results here:

\begin{itemize}
\item{The MWA detection threshold is 10\% less sensitive than our theoretical prediction}
\item{There is a 5\% flattening in the mean of the spectral index distribution from -3.00 to -2.85 between 1382 MHz and 192 MHz, the spectral index distribution in this analysis has width of 0.6, but this has not been constrained by the fits}
\item{This distribution correctly predicts the number of coincident detections and the 50\% correlation between the two observations}
\end{itemize}

The {\em normal} emission from pulsars typically obeys a simple power law of index $-1.8$ \citep{maron00}. However, the MP and IP components of the Crab pulsar have much steeper spectral indices of -2.8 and -3.7 when measured between 410 and 1660 MHz \citep{Manchester71}.  A flattening in the spectra of normal pulsar emission is commonly observed and many young pulsars have been found to display complex spectra and low frequency ($\sim$100\,MHz) spectral turn-overs \citep{malofeev:1994, kijak:2007}. The phase averaged emission from the Crab pulsar is complicated due to the multi-component structure. We are unaware of any observed flattening of MP component at lower frequencies but suspect that this is confused by the rapid evolution of precursor. The cause of the spectral turnover in these objects is not known and experiments are complicated by the difficulty of observing complex pulse profiles at low frequencies.

\subsection{Propagation effects within the nebula}

Our measurements of the scattering timescale (\S \ref{scatt}) indicate that the environment of the nebula continues to affect the observations of the pulsar, as already observed by \citet{Rankin73}, \citet{Isaacman77},  \citet{Bhat08}, \citet{Kuzmin08} and \citet{Ellingson13}. Free-free absorption within the filamentary structures, evident in observations of the Crab nebula, has already been reported by \citet{Bridle} and \citet{Bietenholz97}. We now investigate whether this effect could be responsible for the modest flattening in spectral index that has been indicated by our analysis.
The free-free absorption coefficient ($\alpha^{\rm ff}_{\nu}$) in the Rayleigh-Jeans regime can be numerically approximated (in cgs units) to be:
\begin{equation}
\alpha^{\rm ff}_{\nu} = 0.018T^{-3/2}Z^{2}n_{\rm e}n_{\rm i}\nu^{-2}\bar{g}_{\rm ff}
\end{equation}
where $Z$ is the atomic number of the absorber, $n_{\rm e}$ and $n_{\rm i}$, are the number density per cm$^-3$ of the electrons and ions respectively, and T is the temperature in K \citep{Rybicki79}. $\bar{g}_{\rm ff}$ is the Gaunt factor, which in this regime (the high temperature or {\em quantum limit} ) is approximated by \citet{Gayet70}:
\begin{equation}
\bar{g}_{\rm ff} = \frac{\sqrt{3}}{\pi}\mathrm{ln} \left( \frac{4kT}{h\nu e^{\gamma}}\right),
\end{equation}
where {\it k} is Boltzmann's constant, {\it h} is Planck's constant, and $\gamma$ is Euler's constant. At these temperatures and frequencies the Gaunt factor is approximately 10. The optical depth $\tau_{\rm ff}$ is given by:
\begin{equation}
\tau^{\rm ff}_{\nu} = \int_0^{z(\mathrm{pc})} \alpha^{\rm ff}_{\nu}\, \mathrm{d}z,
\end{equation}
where $z$ is the path length, and the absorption factor is subsequently calculated as $e^{-\tau^{\rm ff}}$.

The nebula has an approximate radius of 1.7 pc and if all the observed DM (56.70 cm$^{-3}$pc) were due to the nebula, then $n_{\rm e}$ would be 56.7/1.7 or 33.4 cm$^{-3}$. Assuming the nebula is predominantly Hydrogen, $Z=1$, $n_{\rm e}=n_{\rm i}=33.4$ cm$^{-3}$ at a temperature 10$^4$K, then at our observing frequency ($192 \times 10^{6}$ Hz) the free-free absorption coefficient is 
$\sim 5.4\times10^{-23}$ cm$^{-1}$. This is insufficient to account for any absorption even at these very high Gaunt factors. 

The ionised material is not evenly distributed, and this inhomogeneity will strongly influence the level of free-free absorption. The filamentary structures of the nebula that have been observed to display free-free absorption in spectral index maps by \citet{Bietenholz97}, have been estimated to have an electron density in excess of 200 cm$^{-3}$. These are arcsecond size structures, implying a size scale of 0.02 pc. Each of these filaments would contribute  $1.2\times10^{-4}$ to the optical  depth and attenuate the intensity by a factor of 0.9998, which also cannot supply the necessary absorption even if several hundred filaments intercept the line of sight. \citet{Davidson70} use line intensities to estimate the filamentary $n_{\rm e}$ to be $\sim 1000$ cm$^{-3}$, however each filament now contributes $\sim$ 10 cm$^{-3}$pc to the dispersion measure so only a small number of crossings are permitted, therefore these structures cannot provide the level of absorption required. 

There is also evidence of very small (AU) scale filamentary structures with electron densities in excess of 10$^4$ cm$^{-3}$ \citep{Smith11}.  If many tens of these small filaments are intercepted, they are capable of generating the few percent level absorption required (few \%). They would also not contribute more than 10 cm$^{-3}$pc to the pulsar dispersion measure. However it is unlikely that these structures exist in sufficient numbers to contribute at this level \citep{Smith11}.

In summary it is unlikely that free-free absorption within the filamentary nebula is capable of attenuating the pulse intensity sufficiently to explain our predictions, and therefore any variation in the spectra of the giant pulses is unlikely to be propagative in nature.

\section{CONCLUSIONS}
In this article, we report observations performed simultaneously with the MWA, at a frequency of 193 MHz (15 MHz bandwidth) and the Parkes radio telescope, at a frequency of 1382 MHz (340 MHz bandwidth).  We detected a total of 55 giant pulses at the MWA, and estimate a power-law index of  $\beta = -3.35 \pm 0.35 $ for their pulse fluence distribution. 23 of the pulses were detected at both the MWA and at Parkes radio telescope. These results are consistent with the spectral index of the giant pulses being drawn from normal distribution, the parameters of which are frequency dependent. The mean of the spectral index distribution varies from $-3.00$ to $-2.85$ between 1382 MHz and 192 MHz. It is unlikely that this flattening can be caused by any propagative effects within the nebula.

In this work we have proposed that the less than total correlation observed is simply a function of the spread in giant pulse spectral indices and the relative sensitivities of the two instruments, but have not proposed a mechanism by which this large range of spectral indices may be generated. Giant pulses are thought to comprise complex temporal substructures on nano-second to micro-second timescales \citep{Han03}. \citet{Eilek02} have noted that these substructures are strongly correlated at two observing frequencies with a small fractional bandwidth, and less correlated as the fractional bandwidth increases, suggesting a complex behaviour as a function of frequency, cannot be ruled out by the behaviour we see in this experiment.

The MWA's large frequency coverage and flexible system design offer unique opportunities to further investigate the wide-band properties of giant pulses. With the full-bandwidth VCS recording now feasible \citep{Tremblay15}, it is possible to perform sensitive observations  simultaneously at multiple frequencies within the 80 - 300 MHz range, by suitably spreading out the 30.72 MHz observing band across a large range in frequency. This will allow us to conduct observations that span large fractional bandwidths, with the prospects of determining the emission bandwidth and spectral nature of giant pulses at low frequencies. 

Combining all the MWA tiles coherently, we will realise a factor of $\sim$11 improvement in sensitivity over this experiment. With an increase in gain by this factor, as well as increase in bandwidth by a factor of 2, the instrument is expected to detect about 10,000 pulses in a one hour observation. This also implies that about 600 pulses are expected to be detectable in a 1.28 MHz subband of the MWA. This increase in sensitivity means that a higher percentage of coincident pulses will be detectable when observing in the coherent mode with the MWA full bandwidth.

\bigskip

\noindent
{\it Acknowledgments:} 
We would like to thank Haydon Knight for useful contributions during the preparation of this paper.
This scientific work makes use of the Murchison Radio-astronomy Observatory, operated by CSIRO. We acknowledge the Wajarri Yamatji people as the traditional owners of the Observatory site. Support for the MWA comes from the U.S. National Science Foundation (grants AST-0457585, PHY-0835713, CAREER-0847753, and AST-0908884), the Australian Research Council (LIEF grants LE0775621 and LE0882938), the U.S. Air Force Office of Scientific Research (grant FA9550-0510247), and the Centre for All-sky Astrophysics (an Australian Research Council Centre of Excellence funded by grant CE110001020). Support is also provided by the Smithsonian Astrophysical Observatory, the MIT School of Science, the Raman Research Institute, the Australian National University, and the Victoria University of Wellington (via grant MED-E1799 from the New Zealand Ministry of Economic Development and an IBM Shared University Research Grant). The Australian Federal government provides additional support via the Commonwealth Scientific and Industrial Research Organisation (CSIRO), National Collaborative Research Infrastructure Strategy, Education Investment Fund, and the Australia India Strategic Research Fund, and Astronomy Australia Limited, under contract to Curtin University. We acknowledge the iVEC Petabyte Data Store, the Initiative in Innovative Computing and the CUDA Center for Excellence sponsored by NVIDIA at Harvard University, and the International Centre for Radio Astronomy Research (ICRAR), a Joint Venture of Curtin University and The University of Western Australia, funded by the Western Australian State government. This research was conducted by the Australian Research Council Centre of Excellence for All-sky Astrophysics (CAASTRO), through project number CE110001020.


\begin{thebibliography}{}
\bibitem[Argyle \& Gower(1972)]{Argyle72} Argyle, E., \& Gower, J.~F.~R.\ 1972, \apjl, 175, L89 
\bibitem[Bhat et al.(2003)]{Bhat03} Bhat, N.~D.~R., Cordes, J.~M., \& Chatterjee, S.\ 2003, \apj, 584, 782
\bibitem[Bhat et al.(2007)]{Bhat07} Bhat, N. D. R., Wayth, R.~B., Knight, H.~S., et al. 2007, \apj, 665, 618-627
\bibitem[Bhat et al.(2008)]{Bhat08} Bhat, N.~D.~R., Tingay, S.~J., \& Knight, H.~S.\ 2008, \apj, 676, 1200 
\bibitem[Bietenholz et al.(1997)]{Bietenholz97} Bietenholz, M.~F., Kassim, N., Frail, D.~A., et al. 1997, \apj, 490, 291 
\bibitem[Bilous et al.(2011)]{Bilous11} Bilous, A.~V., Kondratiev, V.~I., McLaughlin, M.~A., et al.\ 2011, \apj, 728, 110 
\bibitem[Bowman et al.(2013)]{Bowman} Bowman, J.~D., Cairns, I., Kaplan, D.~L., et al.\ 2013, \pasa, 30, e031 
\bibitem[Bridle(1970)]{Bridle} Bridle, A.~H.\ 1970, \nat, 225, 1035
\bibitem[Burke-Spolaor et al.(2011)]{Buke-Spolaor2011} Burke-Spolaor, S., Bailes, M., Johnston, S., et al.\ 2011, \mnras, 416, 2465
\bibitem[Burke-Spolaor et al.(2012)]{Buke-Spolaor2012} Burke-Spolaor, S., Johnston, S., Bailes, M., et al.\ 2012, \mnras, 423, 1351
\bibitem[Comella et al.(1969)]{Comella} Comella, J.~M., Craft, H.~D., Lovelace, R.~V.~E., \& Sutton, J.~M.\ 1969, \nat, 221, 453 
\bibitem[Cordes et al.(2004)]{cordes04} Cordes, J. M., Bhat, N.~D.~R., Hankins, T.~H., McLaughlin, M.~A., \& Kern, J.\ 2004, \apj, 612, 375-388
\bibitem[Davidson \& Tucker(1970)]{Davidson70} Davidson, K., \& Tucker, W.\ 1970, \apj, 161, 437 
\bibitem[Eilek et al.(2002)]{Eilek02} Eilek, J.~A., Arendt, P.~N., Jr., Hankins, T.~H., \& Weatherall, J.~C.\ 2002, Neutron Stars, Pulsars, and Supernova Remnants, ed. W. Becker, H. Lesch, \& J. Tr$\ddot{\mathrm{u}}$mper, 249 
\bibitem[Ellingson et al. (2013)]{Ellingson13}Ellingson, S.~W., Clarke, T.~E., Craig J., et al. 2013, \apj, 786, 136
\bibitem[Ershov \& Kuzmin(2005)]{Ershov05} Ershov, A. A., \& Kuzmin, A. D. 2005, \aap, 443, 593-597
\bibitem[Gabor (1946)]{Gabor} Gabor, D. 1946, Journal of the Institute of Electrical Engineering, 93, 429–457.
\bibitem[Gayet(1970)]{Gayet70} Gayet, R.\ 1970, \aap, 9, 312
\bibitem[Goldstein, Morris \& Yen (2004)]{gmy04} Goldstein, M. L., Morris, S. A., \& Yen, G. G.\ 2004, The European Physical Journal B, 41, 255-258
\bibitem[Hankins et al.(2003)]{Han03} Hankins, T. H., Kern, J.~S., Weatherall, J.~C., \& Eilek, J.~A.\ 2003, \nat, 422, 141-143
\bibitem[Hankins \& Eilek(2007)]{Han07} Hankins, T.~H., \& Eilek, J.~A.\ 2007, \apj, 670, 693 
\bibitem[Hesse \& Wielebinski(1974)]{Hesse74} Hesse, K.~H., \& Wielebinski, R.\ 1974, \aap, 31, 409 
\bibitem[Hobbs et al.(2006)]{Hobbs06} Hobbs, G.~B., Edwards, R.~T., \& Manchester, R.~N.\ 2006, \mnras, 369, 655
\bibitem[Hotan et al.(2004)]{Hotan04} Hotan, A.~W., van Straten, W., \& Manchester, R.~N.\ 2004, \pasa, 21, 302 
\bibitem[Isaacman \& Rankin(1977)]{Isaacman77} Isaacman, R., \& Rankin, J.~M.\ 1977, \apj, 214, 214 
\bibitem[Karuppusamy et al.(2010)]{Karuppusamy10} Karuppusamy, R., Stappers, B.~W., \& van Straten, W.\ 2010, \aap, 515, A36 
\bibitem[Karuppusamy et al.(2012)]{Karuppusamy12} Karuppusamy, R., Stappers, B.~W., \& Lee, K.~J.\ 2012, \aap, 538, A7 
\bibitem[Keith et al.(2010)]{Keith10} Keith, M. J., Jameson, A., van Straten, W., et al. 2010, \mnras, 409, 619-627
\bibitem[Knight et al.(2005)]{Knight05} Knight, H. S., Bailes, M., Manchester, R.~N., \& Ord, S.~M.\ 2005, \apj, 625, 951-956
\bibitem[Kijak et al. (2007)]{kijak:2007} Kijak, J., Gupta, Y., \& Krzeszowski, K. 2007, \aap, 462, 699
\bibitem[Kostyuk et al.(2003)]{Kost03} Kostyuk, S.~V., Kondratiev, V.~I., Kuzmin, A.~D., Popov, M.~V., \& Soglasnov, V.~A.\ 2003, Astronomy Letters, 29, 387 
\bibitem[Kuz'min et al.(2008)]{Kuzmin08} Kuz'min, A.~D., Losovskii, B.~Y., Logvinenko, S.~V., \& Litvinov, I.~I.\ 2008, Astronomy Reports, 52, 910 
\bibitem[Lonsdale et al.(2009)]{Lonsdale09} Lonsdale, C.~J., Cappallo, R.~J., Morales, M.~F., et al.\ 2009, IEEE Proceedings, 97, 1497-1506 
%\bibitem[Lorimer \& Kramer(2005)]{Lorimer05} Lorimer, D., \& Kramer, M.\ 2005, Handbook of Pulsar Astronomy (Cambridge University Press)
\bibitem[Lundgren et al.(1995)]{Lund95} Lundgren, S. C., Cordes, J.~M., Ulmer, M., et al. 1995, \apj, 453, 433	
\bibitem[Lyne \& Rickett(1968)]{Lyne68} Lyne, A.~G., \& Rickett, B.~J.\ 1968, \nat, 219, 1339 
\bibitem[Lyne et al.(1993)]{Lyne93} Lyne, A.~G., Pritchard, R.~S., \& Graham-Smith, F.\ 1993, \mnras, 265, 1003 ( http://www.jb.man.ac.uk/~pulsar/crab.html) 
\bibitem[Malofeev et al.(1994)]{malofeev:1994} Malofeev, V.~M., Gil, J.~A., Jessner, A., et al. 1994, \aap, 285, 201
\bibitem[Manchester et al.(1971)]{Manchester71} Manchester, R.~N., Tademaru, E., \& Smith, F.~G.\ 1971, \nat, 234, 164
\bibitem[Maron et al.(2000)]{maron00} Maron, O., Kijak, J., Kramer, M., \& Wielebinski, R. 2000, \aaps, 147, 195
\bibitem[Mickaliger et al.(2012)]{Mickaliger12} Mickaliger, M.~B., McLaughlin, M.~A., Lorimer, D.~R., et al.\ 2012, \apj, 760, 64 
\bibitem[de Oliveira-Costa et al.(2008)]{Oliveira} de Oliveira-Costa, A., Tegmark, M., Gaensler, B.~M., et al.\ 2008, \mnras, 388, 247 
\bibitem[Popov et al.(2006)] {Pop06} Popov, M.~V., Kuzmin, A.~D., Ulyanov, O.~M., et al. 2006, Instant Radio Spectra of Giant Pulses from the Crab Pulsar Over Decimeter to Decameter Wave Band, IAU, Joint Discussion 2, 19 
\bibitem[Popov \& Stappers(2007)]{Pop07} Popov, M.~V., \& Stappers, B.\ 2007, \aap, 470, 1003 
\bibitem[Rankin \& Counselman(1973)]{Rankin73} Rankin, J.~M., \& Counselman, C.~C., III 1973, \apj, 181, 875 
\bibitem[Ransom(2001)]{Ransom01} Ransom, S.~M.\ 2001, Ph.D.~Thesis
\bibitem[Rickett \& Lyne(1990)]{Rickett90} Rickett, B.~J., \& Lyne, A.~G.\ 1990, \mnras, 244, 68 
\bibitem[Ritchings(1976)]{Ritchings76} Ritchings, R.~T.\ 1976, \mnras, 176, 249 
\bibitem[Romani \& Johnston (2001)]{Rom01} Romani, R. W., \& Johnston, S. 2001, \apjl, 557, 93-96	
\bibitem[Rybicki \& Lightman(1979)]{Rybicki79} Rybicki, G.~B., \& Lightman, A.~P.\ 1979, Astronomy Quarterly, 3, 199
\bibitem[Sallmen et al.(1999)]{Sal99} Sallmen, S., Backer, D.~C., Hankins, T.~H., Moffett, D., \& Lundgren, S. 1999, \apj 517, 460-471
\bibitem[Smith \& Terry(2011)]{Smith11} Smith, K.~W., \& Terry, P.~W.\ 2011, \apj, 730, 133 
\bibitem[Soglasnov et al.(2004)]{so04} Soglasnov, V.~A., Popov, M.~V., Bartel, N., et al.\ 2004, \apj, 616, 439 
\bibitem[Staelin \& Reifenstein(1968)]{sta68} Staelin, D. H., \& Reifestein, E. C.  1968, Science, 162, 1481
\bibitem[Stappers et al.(2011)]{Stappers11} Stappers, B., Hessels, J., Alexov, A., et al.\ 2011, Pulsars and Fast Transients with LOFAR, ed. M. Burgay, N. D'Amico, P. Esposito, A. Pellizzoni \& A. Possenti, p. 325 
\bibitem[Tingay et al.(2013)]{Tingay13} Tingay, S.~J., Goeke, R., Bowman, J.~D., et al. 2013, PASA, 30, 7
\bibitem[Tremblay et al.(2015)]{Tremblay15} Tremblay, S.~E., Ord, S.~M., Bhat, N.~D.~R., et al.\ 2015, \pasa, 32, e005 
\bibitem[Wolszczan et al.(1984)]{Wol84} Wolszczan, A., Cordes, J., \& Stinebring, D. 1984, Birth and Evolution of Neutron Stars: Issues Raised by Millisecond 	Pulsars, ed. S. P. Reinolds \& D. R. Stinebring (NRAO, Green Bank, 1984), p. 63. 	
\end{thebibliography}
\end{document}